\documentclass{article}
\usepackage{spconf,amsmath,graphicx,hyperref}
\usepackage{amsmath}    
\usepackage{amssymb}    
\usepackage{graphicx}   
\usepackage{caption}    
\usepackage{hyperref}   
\usepackage{booktabs}   
\usepackage{multirow}   
\usepackage{cite}       
\usepackage{tabularx}

\title{Can Hierarchical Cross-Modal Fusion Predict Human Perception of AI Dubbed Content?}
\name{Ashwini Dasare, Nirmesh Shah, Ashishkumar Gudmalwar, Pankaj Wasnik \vspace{-0.4cm}}
\address{Sony Research India \vspace{0.2cm} \\ \{ashwini.dasare, nirmesh.shah, ashish.gudmalwar1, pankaj.wasnik\}@sony.com}
\begin{document}
%
\maketitle
\begin{abstract}
Evaluating AI generated dubbed content is inherently multi-dimensional, shaped by synchronization, intelligibility, speaker consistency, emotional alignment, and semantic context. Human Mean Opinion Scores (MOS) remain the gold standard but are costly and impractical at scale. We present a hierarchical multimodal architecture for perceptually meaningful dubbing evaluation, integrating complementary cues from audio, video, and text. The model captures fine-grained features such as speaker identity, prosody, and content from audio, facial expressions and scene-level cues from video and semantic context from text, which are progressively fused through intra and inter-modal layers. Lightweight LoRA adapters enable parameter-efficient fine-tuning across modalities. To overcome limited subjective labels, we derive proxy MOS by aggregating objective metrics with weights optimized via active learning. The proposed architecture was trained on 12k Hindi–English bidirectional dubbed clips, followed by fine-tuning with human MOS. Our approach achieves strong perceptual alignment (PCC $>$ 0.75), providing a scalable solution for automatic evaluation of AI-dubbed content.
\end{abstract}
\section{Introduction}
\label{sec:intro}
AI-based dubbing has advanced rapidly with progress in neural machine translation (NMT), text-to-speech (TTS), and audio-visual (AV) synchronization \cite{wu2023videodubber,sahipjohn2024dubwise}. Despite these developments, assessing the quality of dubbed content remains an open problem \cite{spiteri2025rethinking}. Current evaluation methods focus on isolated dimensions such as, speech naturalness, intelligibility, or AV synchrony, but these do not capture the overall perceptual quality\cite{wang2025towards}. In practice, human evaluators judge dubbed content in a multifaceted way \cite{atilgan2018integration,akhtar2017audio,Gao2022_AudiovisualMeta}, simultaneously considering prosody, speaker identity, semantic consistency, emotional congruence, and temporal alignment. The overall quality of AI-dubbed content from this holistic perspective is still not well characterized, making it difficult to compare systems or optimize toward natural and coherent dubbing\cite{bernabo2025and}.

A key challenge is the lack of scalable human ratings. Mean Opinion Scores (MOS) remain the gold standard, but collecting them is costly, time-consuming, and infeasible at the scale required for modern dubbing systems \cite{bailly2025hot}. To overcome this bottleneck, we introduce Proxy MOS, a weak-supervision strategy that aggregates multiple objective metrics into pseudo-perceptual labels. Unlike simple averaging, weights are adaptively learned through an active learning scheme guided by a small subset of human ratings, improving alignment between proxy labels and human perception.

To model the multifaceted nature of dubbing quality, we design a hierarchical multimodal architecture that integrates audio, video, and text features through progressive intra and inter-modal fusion. Fine-grained cues from audio (speaker identity, prosody), text (semantic intent), and video (character and affective context) jointly influence perceived dubbing quality, motivating multimodal modeling. Pre-trained models serve as encoders for each modality, and lightweight LoRA adapters are employed to project representations into a shared space, enabling parameter efficient adaptation across modalities. Intra-modal fusion aligns complementary cues within each modality, while inter-modal fusion captures dependencies across modalities, ensuring a holistic representation of dubbing quality.

In this paper, we propose a two-stage training pipeline. In the first stage, the model is pre-trained on Proxy MOS labels derived from over 12,000 dubbed clips, generated using state-of-the-art NMT, TTS, and audio-visual synchronization pipelines applied to the publicly available MELD\cite{poria2019meld} and M2H2\cite{chauhan2021m2h2} datasets. In the second stage, the model is fine-tuned on a smaller set of human MOS ratings to refine perceptual alignment. Our contributions are summarized as follows:
\begin{itemize}
    \item Propose a hierarchical multimodal framework that fuses audio, visual, and textual features through intra and inter-modal fusion layers for automatic perceptual evaluation of AI dubbed content.
    \item Incorporate lightweight LoRA adapter layers for parameter-efficient adaptation across modalities.
    \item Propose two-stage training pipeline (1) an active learning–based weak supervision framework that derives Proxy MOS from multiple objective metrics and (2) finetuning by human-rated MOS for stronger perceptual alignment.
\end{itemize}
\begin{figure}[h]
\begin{minipage}[b]{0.9\linewidth}
  \centering
\centerline{\includegraphics[width=9.6cm]{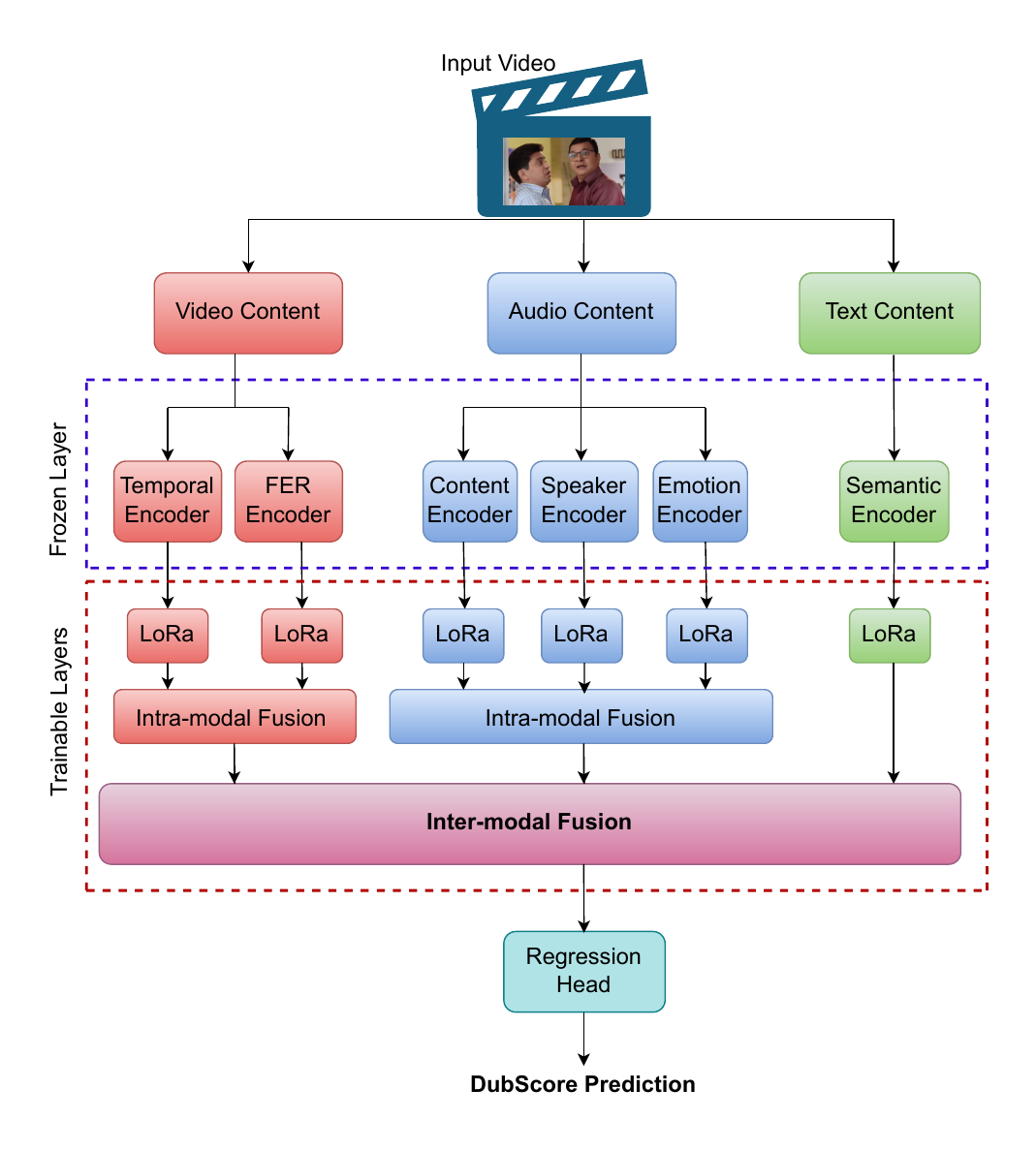}}
  \label{fig:figure1}
\end{minipage}
\vspace{-0.5cm}
\caption{Block diagram of Proposed hierarchical cross-modal fusion architecture.}
\label{fig:archi}
\vspace{-0.5cm}
\end{figure}
\section{Proposed Method}
\subsection{Propose Hierarchical Multimodal Architecture}
\label{sec:architecture}
We propose a hierarchical multimodal network designed to capture both modality-specific cues and cross-modal dependencies critical for perceptual dubbing evaluation, as shown in Figure \ref{fig:archi}. The architecture leverages state-of-the-art pretrained encoders for audio, video, and text, each selected for their demonstrated ability to extract semantically and perceptually rich representations. For video encoders, we extract visual embeddings using the TimeSformer transformer-based video representation model \cite{Bertasius2021_Timesformer}, which captures long-range spatio-temporal dependencies through divided space-time attention. The output embeddings are 768-dimensional. In addition, to incorporate facial affective cues, we also employ a Facial Expression Recognition (FER) encoder \cite{Deng2019_FaceNetFER, Li2023_FER}, which maps cropped face sequences into 512-dimensional emotion-related features. For audio encoders, we use Wav2Vec2.0 \cite{Baevski2020_Wav2Vec2}, which outputs 768-dimensional frame-level embeddings that are temporally pooled and linearly projected to 256 dimensions before fusion. To capture speaker traits and paralinguistic cues, we adopt ECAPA-TDNN \cite{Desplanques2020_ECAPA}, yielding 192-dimensional speaker embeddings. Moreover, we include Emo2Vec \cite{ma2024emotion2vec}, a deep emotional embedding extractor for speech, producing 256-dimensional representations. Finally, for the text encoder, we use the Sentence-BERT encoder \cite{Reimers2019_SentenceBERT}, producing 768-dimensional embeddings that capture semantic similarity at the sentence level. To efficiently adapt these encoders, we employ Low-Rank Adaptation (LoRA) \cite{Hu2021_LoRA}, which introduces lightweight trainable matrices into attention and projection layers while keeping pretrained weights frozen. This reduces the number of trainable parameters while maintaining flexibility.

The hierarchical design reflects how human evaluators perceive dubbed content, first by consolidating heterogeneous cues within each modality, and then integrating them across modalities. For instance, audio encoders provide speech content, speaker identity, and prosodic cues, while video encoders capture both global context and fine-grained facial expressions. Fusing these directly risks information loss or modality dominance, hence the need for staged fusion.
Formally, given embeddings $\{\mathbf{h}^{(i)}_m\}_{i=1}^{N_m}$ for modality 
$m \in \{\text{audio}, \text{video}, \text{text}\}$, we adapt them via: \vspace{-0.1cm}
\begin{equation}
\tilde{\mathbf{h}}^{(i)}_m = P_m(\mathbf{h}^{(i)}_m) + A_m(\mathbf{h}^{(i)}_m),
\vspace{-0.2cm}
\end{equation}
where $P_m$ is a linear projection and $A_m$ is a LoRA adapter. \textbf{Intra-modal fusion} aggregates modality-specific cues using attention-based gating, \vspace{-0.2cm}
\begin{equation}
\mathbf{z}_m = \sum_{i=1}^{N_m} \alpha^{(i)}_m \tilde{\mathbf{h}}^{(i)}_m, 
\quad 
\alpha^{(i)}_m = \frac{\exp(\mathbf{w}^\top \tilde{\mathbf{h}}^{(i)}_m)}{\sum_j \exp(\mathbf{w}^\top \tilde{\mathbf{h}}^{(j)}_m)}.
\vspace{-0.2cm}
\end{equation}
\textbf{Inter-modal fusion:} To account for modality reliability, we apply gating before cross-modal modeling: \vspace{-0.2cm}
\begin{equation}
\hat{\mathbf{z}}_m = g_m(\mathbf{z}_m)\,\mathbf{z}_m, 
\quad g_m(\mathbf{z}_m) = \frac{\exp(\phi(\mathbf{z}_m))}{\sum_{m'} \exp(\phi(\mathbf{z}_{m'}))}.
\vspace{-0.2cm}
\end{equation}
The gated modality-level vectors are concatenated and processed by a 3-layer, 4-head transformer encoder: \vspace{-0.1cm}
\begin{equation}
\mathbf{H} = \text{TransformerEncoder}([\hat{\mathbf{z}}_{\text{audio}};\hat{\mathbf{z}}_{\text{video}};\hat{\mathbf{z}}_{\text{text}}]).
\vspace{-0.2cm}
\end{equation}

Finally, a regression head predicts a single perceptual dubbing score (DubScore) using an L2 (MSE) loss. This hierarchical strategy consolidates intra-modal cues before cross-modal fusion, ensuring both modality-specific refinement and robust interaction aligned with perceptual judgments.

\subsection{Active Learning for Proxy MOS}
Human MOS annotations are limited and costly to obtain, making it infeasible to label the entire dataset. 
To overcome this, we generate Proxy MOS labels by aggregating multiple objective metrics, including 
PEAVS~\cite{Goncalves2024_PEAVS} for audio–visual synchrony, EmoSync~\cite{ryumina2024avcer} for visual emotion alignment, 
LogF0RMSE for speaker consistency, UTMOS~\cite{saeki2022utmos} for speech quality, and SpeechBERT Score~\cite{saeki2024speechbertscore} 
for semantic alignment. The aggregated Proxy MOS is defined as: \vspace{-0.2cm}
\begin{equation}
\text{Proxy MOS} = \sum_{i} w_i \cdot O_i,
\vspace{-0.2cm}
\end{equation}
where $O_i$ denotes the $i$-th objective metric and $w_i$ its learnable weight. The challenge is to estimate $w_i$ using only a limited set of human MOS labels. To this end, we employ 
an active learning (AL) strategy that incrementally refines the weights in three stages. Importantly, 
the reported percentages (33\%, 66\%, 100\%) refer to fractions of the \emph{labeled training subset}, not the 
entire corpus, consistent with standard AL practice where human annotations are scarce.

\textbf{Stage I: Initialization.} A small labeled subset (33\% of available MOS annotations) is used to 
estimate initial weights by maximizing the correlation between Proxy MOS and human MOS. This provides a 
first approximation of perceptual alignment.

\textbf{Stage II: Uncertainty- and Diversity-Based Sampling.} To expand the labeled pool, we query the most informative samples from the remaining unlabeled subset. Samples are ranked by prediction uncertainty (entropy of Proxy MOS) and then filtered to maximize diversity across dubbing conditions (e.g., clip length, 
acoustic background, speaker identity). This hybrid uncertainty–diversity criterion ensures that annotations are 
both informative and representative. An additional 33\% of labeled data is added in this stage (total 66\%).

\textbf{Stage III: Refinement.} The process is repeated to reach 100\% of the labeled subset. We then optimize 
the weights $w_i$ by maximizing Pearson correlation with the expanded human MOS set: \vspace{-0.2cm}
\begin{equation}
\max_{w} \ \rho \big( \text{ProxyMOS}(w), \ \text{HumanMOS} \big),
\vspace{-0.2cm}
\end{equation}
where $\rho$ denotes Pearson correlation. Proxy-MOS weights are learned by maximizing Pearson correlation with human MOS (Eq. 6), while active learning selects informative clips based on uncertainty and diversity. The learned weights are used to compute Proxy MOS labels for the full 12k dubbed clips, which provide weak supervision for training the hierarchical multimodal network at scale. The model is then fine-tuned on a limited set of human MOS annotations, yielding a perceptually aligned predictor.

\section{Experimental Results and Discussion}
\subsection{Datasets and Experimental Setup}
\label{sec:data}
We perform our evaluation on two publicly available datasets, namely MELD \cite{poria2019meld} and M2H2 \cite{chauhan2021m2h2}. MELD (English) was dubbed into Hindi, and M2H2 (Hindi) into English. Both datasets contain video clips, along with speaker tags, emotion tags and transcripts. For creative translation, we used the Gemini-9B model, prompting it with speaker attributes and emotion tags to preserve both semantic and affective context. Translated utterances were synthesized into the target language using the F5-TTS system for the generation of expressive speech. Finally, to ensure audio-visual synchrony, we applied a global time-stretching algorithm that aligns synthesized audio with the original video. This pipeline produced approximately 6k dubbed clips from MELD and 4k from M2H2, covering diverse speakers and emotions. Additionally, we included 2k original ground-truth clips to stabilize Proxy MOS training, resulting in a total of 12k clips 
Details of pre-trained encoders are provided in Section 2.1. For adaptation, LoRA modules were attached to each encoder with a rank of $r=16$, empirically chosen to balance performance and parameter efficiency. Training was performed using the Adam optimizer with a learning rate of 1e-4, batch size 64, dropout 0.2, for 50 epochs. All results are reported using 4-fold cross-validation to ensure robustness..
\subsection{Subjective Evaluation}
For subjective evaluation, we recruited 30 participants (aged 25–40, no reported hearing or vision impairments). A total of 1350 ratings were collected, with each participant evaluating 45 dubbed video clips. In addition to overall dubbing quality, participants rated six rubric-based aspects aligned with established dubbing evaluation standards \cite{spiteri2024quality}: audio–visual synchrony, speaker consistency, voice clarity, emotional expressiveness, prosody in speech, and overall dubbing quality. 
The human-rated MOS set was divided into an 80\%–20\% train–test split for fine-tuning the proposed network. To further assess reliability of human-rated MOS, we conducted an inter-listener agreement analysis (as shown in Table \ref{tab:ICC}). The results show good internal consistency (Cronbach’s $\alpha = 0.82$). The intraclass correlation coefficients indicate moderate reliability, with ICC1 = 0.69 (individual ratings) and ICC2 = 0.59 (aggregated MOS).
\begin{table}[h]
\vspace{-0.2cm}
\centering
\caption{Inter-listener agreement metrics.}
\vspace{-0.2cm}
\label{tab:ICC}
\begin{tabular}{l c}
\hline
\textbf{Metric} & \textbf{Value} \\
\hline
Cronbach's Alpha & 0.82 \\
ICC1 (One-way random) & 0.69\\ 
ICC2 (Two-way random) & 0.59 \\ 
\hline
\end{tabular}
\vspace{-0.5cm}
\end{table}
\subsection{Experimental Results}
We evaluate models using Pearson’s correlation coefficient (PCC), Spearman’s rank correlation coefficient (SRCC), and mean squared error (MSE) between the predicted DubScore and the human-rated MOS. MOS labels ([1,5]) are normalized during training and rescaled at inference. Table~\ref{tab:multimodal} reports the performance of the proposed network under different modality configurations, including unimodal, bimodal, and the full multi-modal setting. This ablation highlights the individual and complementary contributions of each modality. Among single modalities, audio provides the strongest predictive signal, text offers moderate performance, and video contributes little when used alone, especially in complex scenes with limited visual identity cues. Bimodal fusion improves results, with audio–text outperforming other pairs. The proposed full multi-modal system achieves the best alignment with human perception, confirming that hierarchical integration of all modalities provides complementary benefits beyond unimodal or bimodal systems.\\
\begin{table}[h]
\centering
\caption{Performance comparison of the proposed network under different modality configurations. Here, Audio (A), Video (V) and Text (T) represents considered modalities.}
\vspace{-0.2cm}
\label{tab:multimodal}
\begin{tabular}{ccccc}
\hline
Modality & PCC $\uparrow$ & SRCC $\uparrow$ & MSE $\downarrow$ \\
\hline
A   & 0.68 & 0.60 & 4.30 \\
V   & 0.05 & 0.01 & 3.84 \\
T   & 0.34 & 0.43 & 3.84 \\
A+V  & 0.71 & 0.65 & 3.88 \\
A+T  & 0.73 & 0.76 & 4.39 \\
V+T  & 0.50 & 0.54 & \textbf{3.77} \\
A+V+T & \textbf{0.76} & \textbf{0.77} & 3.88 \\
\hline
\end{tabular}
\vspace{-0.5cm}
\end{table} 

Since Proxy MOS relies on learning weights for objective measures, we employ active learning (AL) to make this process data-efficient. A random-sampling baseline is used for comparison to isolate the benefit of AL. We adopt a three-stage setup, where labeled data is incrementally expanded from 33\% to 100\% of the training set, with the most informative samples selected at each step. Table \ref{tab:uncertainty} shows that calibration steadily improves as label budgets grow. For example, Average Predictive Variance (APV) decreases, Prediction Interval Coverage Probability (PICP) increases, Mean Prediction Interval Width (MPIW) tightens, and Expected Calibration Error (ECE) reduces together indicating more reliable weight estimation. In addition, Table \ref{tab:performance} compares AL with random sampling, showing consistent gains in PCC, SRCC, predictive accuracy (Mean Squared Error (MSE)), and explained variance ($R^2$), with significant improvements at full budget. These results justify the use of AL for Proxy MOS weight learning, which then serves as weak supervision for our hierarchical multimodal network.
\begin{table}[h]
\vspace{-0.1cm}
\centering
\caption{Uncertainty calibration metrics at different label proportions.}
\vspace{-0.2cm}
\label{tab:uncertainty}
\begin{tabular}{lcccc}
\hline
\#Labeled & APV$\downarrow$. & PICP\% $\uparrow$ & MPIW\% $\downarrow$ & ECE $\downarrow$ \\
\hline
33\%  & 0.51 & 78\% & 0.82 & 0.14 \\
66\%  & 0.26 & 82\% & 0.65 & 0.09 \\
100\% & 0.16 & 86\% & 0.58 & 0.06 \\
\hline
\end{tabular}
\vspace{-0.5cm}
\end{table}
\begin{table}[h]
\centering
\caption{Performance comparison of Active Learning (AL) vs. Random (Ra) sampling. 
P-values indicate the statistical significance of improvements of AL over the corresponding Random baseline at the same label budget. Baseline rows (Ra) are not applicable.}
\vspace{-0.2cm}
\label{tab:performance}
\begin{tabular}{l c c c c }
\hline
 Strategy 
& PCC $\uparrow$ & SRCC $\uparrow$ & $R^2$ $\uparrow$ & p-value \\
\hline
Ra (33\%)   & 0.68 & 0.67 & 0.46 & -- \\
AL (S0, 33\%)    & 0.71 & 0.69 & 0.50 & 0.18 \\
Ra(66\%)    & 0.73 & 0.71 & 0.55 & -- \\
AL (S1, 66\%)    & 0.77 & 0.75 & 0.61 & 0.07 \\
Ra (100\%)    & 0.76 & 0.74 & 0.62 & -- \\
AL (S2, 100\%)    & 0.82 & 0.81 & 0.69 & 0.03 \\
\hline
\end{tabular}
\vspace{-0.3cm}
\end{table}



%

We also considered equal-weights strategy as a baseline in proxy mos calculation. Table~\ref{tab:hierarchical_strategies} reports the results. While equal weighting provides a reasonable baseline, AL-based weak supervision consistently yields higher correlations (PCC, SRCC) and lower errors (MSE). Moreover, when combined with fine-tuning on human-rated MOS, AL achieves the overall best performance. These findings confirm that AL-based proposed Proxy MOS is more effective than simple averaging and that its combination with human supervision provides the most reliable perceptual predictions.


\begin{table}[h]
\vspace{-0.1cm}
\centering
\caption{Performance comparison across different training strategies. Here, EW = equal weight, WS = weak supervision with Proxy Mos, and FT = fine-tuning with human MOS.}
\vspace{-0.1cm}
\label{tab:hierarchical_strategies}
\begin{tabular}{p{4cm}ccc} 
\hline
Proxy MOS Strategies & PCC $\uparrow$ & SRCC $\uparrow$ & MSE $\downarrow$ \\
\hline
EW: WS  & 0.22 & 0.25 & 8.14 \\
EW: WS + FT & 0.35 & 0.33 & 5.14 \\
AL: WS & 0.68 & 0.67 & 2.96 \\
AL: WS + FT & \textbf{0.76} & \textbf{0.77} & \textbf{2.70} \\
\hline
\end{tabular}
\vspace{-0.2cm}
\end{table}

\begin{figure}[h]
\vspace{-0.2cm}
\centering
\includegraphics[width=0.9\linewidth]{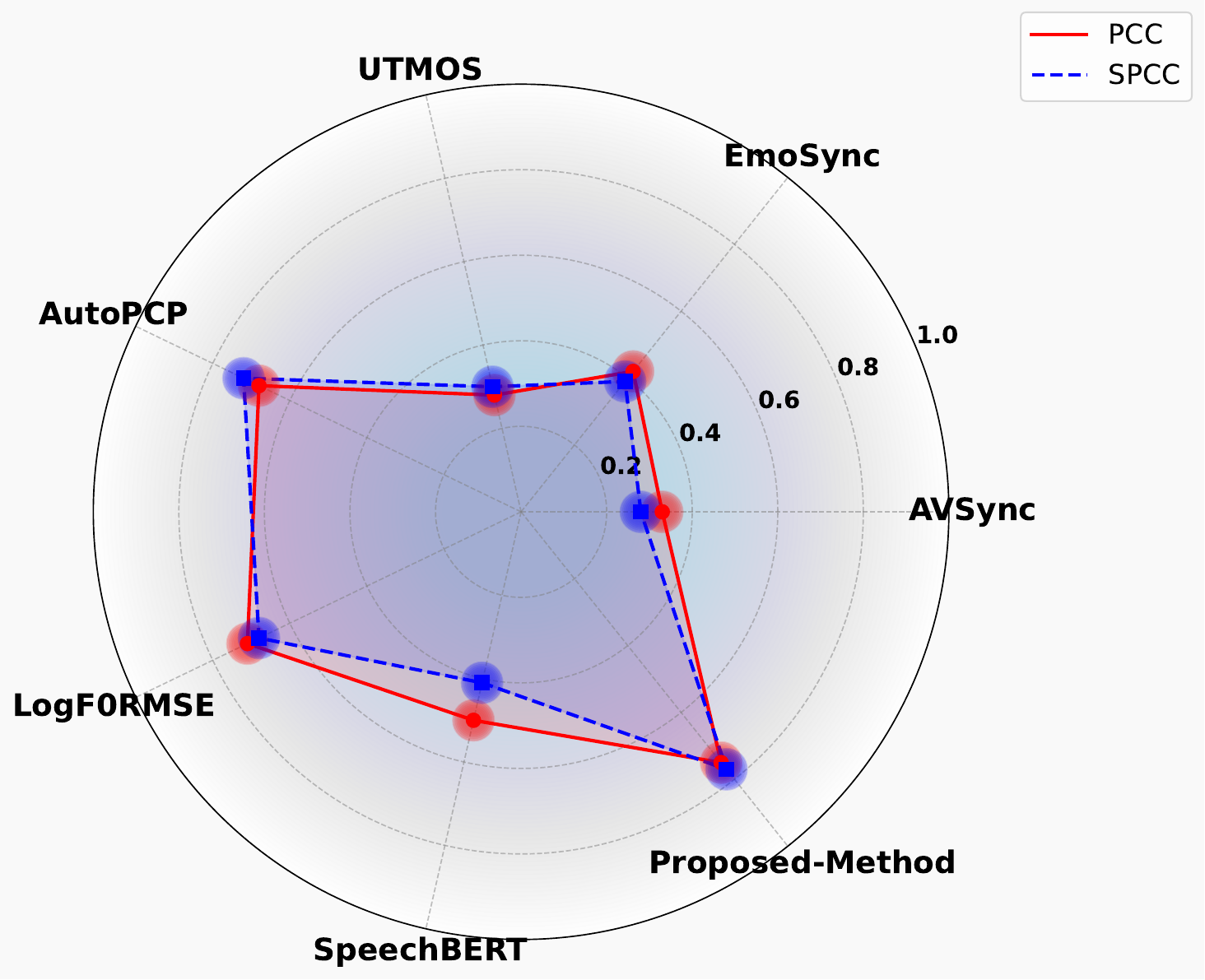}
\label{fig:figure2}
\vspace{-0.2cm}
\caption{Effectiveness in predicting overall dubbing quality scores of different objective metrics}
\vspace{-0.5cm}
\label{fig:spider}
\end{figure}

The radar chart (Fig. \ref{fig:spider}) shows that individual objective metrics yield low PCC and SRCC scores when predicting overall dubbing quality. In contrast, the proposed architecture achieves consistently higher correlations, demonstrating its effectiveness as a unified perceptual evaluation metric.
\section{Conclusion}
This paper presents a hierarchical multimodal architecture for dubbing quality assessment that fuses audio, video, and text cues through intra- and inter-modal layers, achieving strong alignment with human perception. An adaptive active learning strategy with parameter-efficient LoRA fine-tuning enables scalable training using proxy MOS with limited human annotations. Experimental results demonstrate scalable, perceptually aligned dubbing quality prediction and establish a foundation for future work in automated, human-centered audio-visual quality assessment.
\vfill\pagebreak
\bibliographystyle{IEEEbib}
\bibliography{refs}
\end{document}